\documentclass[sigconf,natbib,nonacm]{acmart}
\pdfoutput=1
\usepackage{acronym}
\usepackage[inline]{enumitem}
\usepackage{color}
\usepackage{array}
\usepackage{tikz}
\usepackage{booktabs}
\usepackage{makecell}
\usepackage{soul}
\usepackage{multirow}

\usepackage{pgfplots}
\usepgfplotslibrary{fillbetween}
\pgfplotsset{compat=1.16}

\author{Rolf Jagerman}
\orcid{0000-0002-5169-495X}
\affiliation{
    \institution{Google Research}
    \country{}
    \city{}
}
\email{jagerman@google.com}

\author{Honglei Zhuang}
\orcid{0000-0001-8134-1509}
\affiliation{
    \institution{Google Research}
    \country{}
    \city{}
}
\email{hlz@google.com}

\author{Zhen Qin}
\orcid{0000-0001-6739-134X}
\affiliation{
    \institution{Google Research}
    \country{}
    \city{}
}
\email{zhenqin@google.com}

\author{Xuanhui Wang}
\orcid{0009-0000-1388-1423}
\affiliation{
    \institution{Google Research}
    \country{}
    \city{}
}
\email{xuanhui@google.com}

\author{Michael Bendersky}
\orcid{0000-0002-2941-6240}
\affiliation{
    \institution{Google Research}
    \country{}
    \city{}
}
\email{bemike@google.com}

\copyrightyear{2023}
\acmYear{2023}
\setcopyright{acmlicensed}
\acrodef{LTR}[LTR]{Learning-to-Rank}
\acrodef{IR}[IR]{Information Retrieval}
\acrodef{PRF}[PRF]{Pseudo-Relevance Feedback}
\acrodef{CoT}[CoT]{Chain-of-Thought}
\acrodef{LLM}[LLM]{Large Language Model}
\acrodefplural{LLM}[LLMs]{Large Language Models}

\definecolor{g900grey}{HTML}{202124}
\definecolor{g900blue}{HTML}{174EA6}
\definecolor{g900red}{HTML}{A50E0E}
\definecolor{g900yellow}{HTML}{E37400}
\definecolor{g900green}{HTML}{0D652D}
\definecolor{g700grey}{HTML}{5F6368}
\definecolor{g700blue}{HTML}{1967d2}
\definecolor{g700red}{HTML}{c5221f}
\definecolor{g700yellow}{HTML}{f29900}
\definecolor{g700green}{HTML}{188038}
\definecolor{g500grey}{HTML}{9AA0A6}
\definecolor{g500blue}{HTML}{4285F4}
\definecolor{g500red}{HTML}{EA4335}
\definecolor{g500yellow}{HTML}{FBBC04}
\definecolor{g500green}{HTML}{34A853}
\definecolor{g100grey}{HTML}{F1F3F4}
\definecolor{g100blue}{HTML}{D2E3FC}
\definecolor{g100red}{HTML}{FAD2CF}
\definecolor{g100yellow}{HTML}{FEEFC3}
\definecolor{g100green}{HTML}{CEEAD6}

\newcommand{\todo}[2][0]{{\color{g500red}\ifx#10 \else TODO(#1): \fi#2}}
\newcommand{\sigsym}{{$^\blacktriangle$}~}

\newcommand{\sig}{\rlap{\sigsym}}

\newcommand{\QTD}{{Q2D}}
\newcommand{\QTDZS}{{Q2D/ZS}}
\newcommand{\QTDPRF}{{Q2D/PRF}}
\newcommand{\QTE}{{Q2E}}
\newcommand{\QTEZS}{{Q2E/ZS}}
\newcommand{\QTEPRF}{{Q2E/PRF}}
\newcommand{\CoT}{{CoT}}
\newcommand{\CoTPRF}{{CoT/PRF}}

\newcommand{\query}{\textcolor{g900red}{\{query\}}}
\newcommand{\querya}{\textcolor{g900green}{\{query 1\}}}
\newcommand{\queryb}{\textcolor{g900green}{\{query 2\}}}
\newcommand{\queryc}{\textcolor{g900green}{\{query 3\}}}
\newcommand{\queryd}{\textcolor{g900green}{\{query 4\}}}
\newcommand{\expa}{\textcolor{g900green}{\{expansion 1\}}}
\newcommand{\expb}{\textcolor{g900green}{\{expansion 2\}}}
\newcommand{\expc}{\textcolor{g900green}{\{expansion 3\}}}
\newcommand{\expd}{\textcolor{g900green}{\{expansion 4\}}}
\newcommand{\doca}{\textcolor{g900green}{\{doc 1\}}}
\newcommand{\docb}{\textcolor{g900green}{\{doc 2\}}}
\newcommand{\docc}{\textcolor{g900green}{\{doc 3\}}}
\newcommand{\docd}{\textcolor{g900green}{\{doc 4\}}}
\newcommand{\prfa}{\textcolor{g900green}{\{PRF doc 1\}}}
\newcommand{\prfb}{\textcolor{g900green}{\{PRF doc 2\}}}
\newcommand{\prfc}{\textcolor{g900green}{\{PRF doc 3\}}}

\newcommand{\hlc}[2][yellow]{{%
    \colorlet{foo}{#1}%
    \sethlcolor{foo}\hl{#2}}%
}

\title{Query Expansion by Prompting Large Language Models}

\begin{document}
\begin{abstract}
Query expansion is a widely used technique to improve the recall of search systems.
In this paper, we propose an approach to query expansion that leverages the generative abilities of \acfp{LLM}.
Unlike traditional query expansion approaches such as \ac{PRF} that relies on retrieving a good set of pseudo-relevant documents to expand queries, we rely on the generative and creative abilities of an \ac{LLM} and leverage the knowledge inherent in the model.
We study a variety of different prompts, including zero-shot, few-shot and \acf{CoT}.
We find that \ac{CoT} prompts are especially useful for query expansion as these prompts instruct the model to break queries down step-by-step and can provide a large number of terms related to the original query.
Experimental results on MS-MARCO and BEIR demonstrate that query expansions generated by \acp{LLM} can be more powerful than traditional query expansion methods.
\end{abstract}
\maketitle
\section{Introduction}
Query expansion is a widely used technique that improves the recall of search systems by adding additional terms to the original query.
The expanded query may be able to recover relevant documents that had no lexical overlap with the original query.
Traditional query expansion approaches are typically based on \acf{PRF}~\cite{rocchio1971relevance, robertson1976relevance, robertson1990term, amati2002probabilistic}, which treats the set of retrieved documents from the original query as ``pseudo-relevant'' and uses those documents' contents to extract new query terms.
However, \ac{PRF}-based approaches assume that the top retrieved documents are relevant to the query.
In practice the initial retrieved documents may not be perfectly aligned with the original query, especially if the query is short or ambiguous.
As a result, \ac{PRF}-based approaches may fail if the initial set of retrieved documents is not good enough.

\begin{figure}
\centering
\begin{tikzpicture}[scale=0.82]

\node[draw, minimum height=1.5cm, minimum width=5.5cm, text width=5cm, align=left, rounded corners=5pt]
  (prompt) at (0, 0)
  {Answer the following question:\\\textcolor{g700red}{\textbf{\{query\}}}\\Give the rationale before answering};

\node[draw, minimum height=0.9cm, minimum width=5cm, text width=4cm, align=center, fill=g100blue]
  (llm) at (0, -2)
  {\acf{LLM}};

\node[draw, minimum height=0.7cm, minimum width=5.5cm, text width=5cm, align=center, rounded corners=5pt]
  (expanded) at (0, -3.5)
  {Concat(\textcolor{g700red}{\textbf{\{query\}}}, \textcolor{g700blue}{\textbf{\{model output\}}})};

\node[draw, minimum height=0.9cm, minimum width=5cm, text width=4cm, align=center, fill=g100yellow]
  (bm25) at (0, -5)
  {Retrieval System (BM25)};

\draw[->,>=latex] (prompt) -- (llm);
\draw[->,>=latex] (llm) -- (expanded);
\draw[->,>=latex] (expanded) -- (bm25);
ß
\end{tikzpicture}
\caption{High-level overview of using a zero-shot \acf{CoT} prompt to generate query expansion terms.}
\label{fig:overview}
\Description[A high-level overview of using a CoT prompt to generate query expansions.]{Four boxes on top of each other connected by arrows. The first box contains the prompt "Answer the following question: {question} Give the rationale before answering". The second box represents a Large Language Model (LLM). The third boxes contains the text 'Concat({query}, {model output}). Finally, the fourth box contains the text "Retrieval System (BM25)".}
\end{figure}
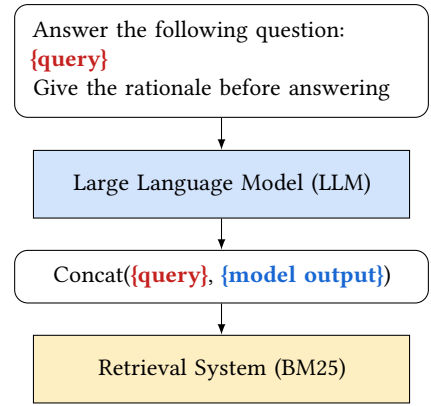

In this paper we propose the use of \acfp{LLM}~\cite{devlin2018bert,raffel2020exploring,brown2020language} to aid in query expansion. \acp{LLM} have seen a growing interest in the \ac{IR} community in recent years.
They exhibit several properties, including the ability to answer questions and generate text, that make them powerful tools.
We propose using those generative abilities to generate useful query expansions.
In particular we investigate ways to prompt an \ac{LLM} and have it generate a variety of alternative and new terms for the original query.
This means that, instead of relying on the knowledge within \ac{PRF} documents or lexical knowledge bases, we rely on the knowledge inherent in the \ac{LLM}.
An example of the proposed methodology is presented in Figure~\ref{fig:overview}.

Our main contributions in this work are as follows:
First, we formulate various prompts to perform query expansion (zero-shot, few-shot and \ac{CoT}) with and without PRF to study their relative performance.
Second, we find that \acf{CoT} prompts perform best and hypothesize that this is because \ac{CoT} prompts instruct the model to break its answer down step-by-step which includes many keywords that can aid in query expansion.
Finally, we study the performance across various model sizes to better understand the practical capabilities and limitations of an \ac{LLM} approach to query expansion.

\section{Related Work}

Query expansion is widely studied~\cite{efthimiadis1996query,carpineto2012survey}. At its core, query expansion helps retrieval systems by expanding query terms into new terms that express the same concept or information need, increasing the likelihood of a lexical match with documents in the corpus. Early works on query expansion focused on either using lexical knowledge bases~\cite{bhogal2007review,voorhees1994query,qiu1993concept} or \acf{PRF}~\cite{robertson1990term,amati2002probabilistic,rocchio1971relevance}.
\ac{PRF}-based approaches are particularly useful in practice because they do not need to construct a domain-specific knowledge base and can be applied to any corpus.
Orthogonal to query expansion is \emph{document expansion}~\cite{tao2006language,efron2012improving,nogueira2019document,zheng2020bert} which applies similar techniques but expands document terms during indexing instead of query terms during retrieval.

Recent works on query expansion have leveraged neural networks to generate or select expansion terms~\cite{imani2019deep,roy2016using,zheng2020bert,zheng2021contextualized}, generally by either training or fine-tuning a model.
In contrast, our work leverages the abilities inherent in \emph{general-purpose} \acp{LLM} without needing to train or fine-tune the model.

We note that our work is similar to the recent works of~\cite{claveau2021neural} and \cite{wang2023query2doc}: leveraging an \ac{LLM} to expand a query.
However, we differentiate our work in several important ways:
First, we study a number of different prompts whereas~\cite{wang2023query2doc} focuses on a single few-shot prompt and~\cite{claveau2021neural} does not study prompts.
Second, unlike~\cite{wang2023query2doc} and~\cite{claveau2021neural}, we focus on generating \emph{query expansion terms} instead of entire pseudo documents. To this end, we demonstrate the performance of our prompts on a variety of \emph{smaller} model sizes which helps understand both the limitations and the practical capabilities of an \ac{LLM} approach to query expansion.
Finally, we experiment with entirely open-source models, inviting reproducibility and openness of research, while~\cite{wang2023query2doc} experiments with a single type of model which is only accessible through a third-party API.

\section{Methodology}
We formulate the query expansion problem as follows: given a query $q$ we wish to generate an \emph{expanded query} $q'$ that contains additional query terms that may help in retrieving relevant documents.
In particular we study the use of an \ac{LLM} to expand the query terms and generate a new query $q'$. Since the \ac{LLM} output may be verbose, we repeat the original query terms 5 times to upweigh their relative importance. This is the same as the trick employed by~\cite{wang2023query2doc}. More formally:
\begin{equation}
q' = \operatorname{Concat}(q, q, q, q, q, \operatorname{LLM}(\textit{prompt}_q)),
\end{equation}
where $\operatorname{Concat}$ is the string concatenation operator, $q$ is the original query, $\operatorname{LLM}$ is a \acl{LLM} and $\textit{prompt}_q$ is the generated prompt based on the query (and potentially side information like few-shot examples or PRF documents).

In this paper we study eight different prompts:
\begin{itemize}[align=parleft,leftmargin=1.6cm,labelsep=0.8cm]
    \item[\textbf{\QTD}] The Query2Doc~\cite{wang2023query2doc} few-shot prompt, asking the model to write a passage that answers the query.
    \item[\textbf{\QTDZS}] A zero-shot version of \textbf{\QTD{}}.
    \item[\textbf{\QTDPRF}] A zero-shot prompt like \textbf{\QTDZS{}} but which also contains extra context in the form of top-3 retrieved \ac{PRF} documents for the query.
    \item[\textbf{\QTE}] Similar to the Query2Doc few-shot prompt but with examples of query \emph{expansion terms} instead of \emph{documents}.
    \item[\textbf{\QTEZS}] A zero-shot version of \textbf{\QTE{}}.
    \item[\textbf{\QTEPRF}] A zero-shot prompt like \textbf{\QTEZS{}} but with extra context in the form of \ac{PRF} documents like \textbf{\QTDPRF{}}.
    \item[\textbf{\CoT}] A zero-shot \acl{CoT} prompt which instructs the model to provide rationale for its answer.
    \item[\textbf{\CoTPRF}] A prompt like \textbf{\CoT{}} but which also contains extra context in the form of top-3 retrieved \ac{PRF} documents for the query.
\end{itemize}
Zero-shot prompts (\textbf{\QTDZS{}} and \textbf{\QTEZS{}}) are the simplest as they consist of a simple plaintext instruction and the input query. Few-shot prompts (\textbf{\QTD{}} and \textbf{\QTE{}}) additionally contain several examples to support in-context learning, for example they contain queries and corresponding expansions. \acl{CoT} (\textbf{\CoT{}}) prompts formulate their instruction to obtain a more verbose output from the model by asking it to break its response down step-by-step. Finally, \acl{PRF} ($\cdot$/\textbf{PRF}) variations of prompts use the top-3 retrieved documents as additional context for the model. See Appendix~\ref{appendix:prompts} for the exact prompts that are used in the experiments.

\section{Experiments}
To validate the effectiveness of the \ac{LLM}-based query expansion we run experiments on two retrieval tasks: MS-MARCO~\cite{nguyen2016ms} passage retrieval and BEIR~\cite{thakur2021beir}.
For the retrieval system we use BM25~\cite{robertson1976relevance,robertson1995okapi} as implemented by Terrier~\cite{ounis2005terrier}\footnote{\url{http://terrier.org/}}.
We use the default BM25 parameters ($b = 0.75$, $k_1 = 1.2$, $k_3 = 8.0$) provided by Terrier.

\subsection{Baselines}
To analyze the \ac{LLM}-based query expansion methods we compare against several classical \ac{PRF}-based query expansion methods~\cite{amati2002probabilistic}:
\begin{itemize}
    \item Bo1: Bose-Einstein 1 weighting
    \item Bo2: Bose-Einstein 2 weighting
    \item KL: Kullback-Leibler weighting
\end{itemize}
The implementations for these are provided by Terrier. In all cases we use the default Terrier settings for query expansion: 3 \ac{PRF} docs and 10 expansion terms.

Furthermore, we include the prompt from Query2Doc~\cite{wang2023query2doc} as a baseline.
However, we do not compare against their exact setup since they use a significantly larger model than the models we study in this paper. The comparisons in this paper are focused on prompts and not on the exact numbers produced by different, potentially much larger, models. Furthermore, for models with a small receptive field (specifically the Flan-T5 models) we only use a 3-shot \QTD{} prompt instead of the standard 4-shot prompt to prevent the prompt from being truncated.

\subsection{Language Models}

We compare the prompts on two types of models, Flan-T5~\cite{raffel2020exploring,chung2022scaling} and Flan-UL2~\cite{tay2022unifying}, at various model sizes:
\begin{itemize}
    \item Flan-T5-Small (60M parameters)
    \item Flan-T5-Base (220M parameters)
    \item Flan-T5-Large (770M parameters)
    \item Flan-T5-XL (3B parameters)
    \item Flan-T5-XXL (11B parameters)
    \item Flan-UL2 (20B parameters)
\end{itemize}

We choose to use the Flan~\cite{wei2021finetuned,chung2022scaling} versions of the T5~\cite{raffel2020exploring} and UL2~\cite{tay2022unifying} models as they are fine-tuned to follow instructions which is critical when using prompt-based approaches.
Furthermore, all of these models are available as open-source\footnote{Models are available at \url{https://huggingface.co/docs/transformers/model_doc/flan-t5} and \url{https://huggingface.co/google/flan-ul2}}.

\subsection{Metrics}
Since we are interested in query expansion, which is largely focussed on improving the recall of first-stage retrieval, we use Recall@1K as our core evaluation metric. We also report top-heavy ranking metrics using MRR@10~\cite{voorhees1999trec} and NDCG@10~\cite{jarvelin2002cumulated} to better understand how the models change the top retrieved results.
We report all our results with significance testing using a paired $t$-test and consider a result significant at $p<0.01$.

\section{Results}

\subsection{MS-MARCO Passage Ranking}

Table~\ref{tbl:msmarco} presents the results on the MS-MARCO passage ranking task. The classical query expansion baselines (Bo1, Bo2 and KL), already provide a useful gain in terms of Recall@1K over the standard BM25 retrieval.
In line with the results of~\cite{harmanrelevance}, we observe that this increase in recall comes at the cost of top-heavy ranking metrics such as MRR@10 and NDCG@10.

Next, we see the results of \ac{LLM}-based query expansion depend heavily on the type of prompts used. 
Similar to the findings of~\cite{wang2023query2doc}, the Query2Doc prompt (\textbf{\QTD{}}) can provide a substantial gain in terms of Recall@1K over the classical approaches.
Interestingly, Query2Doc does not only improve recall, but also improves the top-heavy ranking metrics such as MRR@10 and NDCG@10, providing a good improvement across metrics.
This contrasts with classical query expansion methods which typically sacrifice top-heavy ranking metrics in order to improve recall.

Finally, the best performance is obtained by \textbf{\CoT{}} (and the corresponding PRF-enhanced prompt \textbf{\CoTPRF{}}).
This particular prompt instructs the model to generate a verbose explanation by breaking its answer down into steps.
We hypothesize that this verbosity may lead to many potential keywords that are useful for query expansion.
Finally, we find that adding \ac{PRF} documents to the prompt helps significantly in top-heavy ranking metrics like MRR@10 and NDCG@10 across models and prompts.
A possible explanation for this is that \acp{LLM} are effective in distilling the \ac{PRF} documents, which may already contain relevant passages, by attending over the most promising keywords and using them in the output. We provide a more concrete example of the prompt output in Appendix~\ref{appendix:examples}.

\begin{table}
    \caption{LLM-based query expansion on the MS-MARCO passage ranking dev set. \sigsym indicates a statistically significant (paired $t$-test, $p<0.01$) improvement relative to the \QTD{} Flan-UL2 method. The best result per metric is bolded.}
    \centering
    \small
    \begin{tabular}{lrrr}
    \toprule
    & Recall@1K & MRR@10 & NDCG@10 \\
    \midrule
    BM25 & 87.82 & 18.77 & 23.44 \\
    BM25 + Bo1 & 88.68 & 17.75 & 22.48 \\
    BM25 + Bo2 & 88.32 & 17.58 & 22.30 \\
    BM25 + KL & 88.62 & 17.71 & 22.44 \\
    \midrule
    \textbf{Flan-T5-XXL} (11B) & & & \\
    \quad \QTD{} & 88.76 & 19.07 & 23.76 \\
    \quad \QTDZS{} & 88.88 & 18.55 & 23.13 \\
    \quad \QTDPRF{} & 89.31 & 22.13\sig & 26.43\sig \\
    \quad \QTE{} & 87.74 & 18.74 & 23.37 \\
    \quad \QTEZS{} & 87.93 & 18.79 & 23.45 \\
    \quad \QTEPRF{} & 88.20 & 19.20 & 23.83 \\
    \quad \CoT{} & 89.86 & 19.16 & 23.82 \\
    \quad \CoTPRF{} & 89.02 & 22.08\sig & 26.32\sig \\
    \midrule
    \textbf{Flan-UL2} (20B) & & & \\
    \quad \QTD{} & 89.87 & 19.22 & 23.96 \\
    \quad \QTDZS{} & 86.60 & 15.56 & 19.54 \\
    \quad \QTDPRF{} & 89.28 & 21.42\sig & 25.82\sig \\
    \quad \QTE{} & 88.04 & 18.84 & 23.52 \\
    \quad \QTEZS{} & 88.11 & 18.87 & 23.56 \\
    \quad \QTEPRF{} & 88.43 & 19.24 & 23.90 \\
    \quad \CoT{} & \textbf{90.61}\sig & 20.05\sig & 24.85\sig \\
    \quad \CoTPRF{} & 89.30 & \textbf{22.62}\sig & \textbf{26.89}\sig \\
    \bottomrule
    \end{tabular}
    \label{tbl:msmarco}
\end{table}

\subsection{BEIR}

The BEIR datasets comprise many different zero-shot information retrieval tasks from a variety of domains.
We compare the performance of the different prompts on the BEIR datasets in Table~\ref{tbl:beir}.
The first thing to observe here is that the classical \ac{PRF}-based query expansion baselines still work very well, especially on domain-specific datasets such as trec-covid, scidocs and touche2020. These datasets are largely academic and scientific in nature, and the \ac{PRF} documents may provide useful query terms in these cases. In contrast, the general purpose \acp{LLM} may not have sufficient domain knowledge to be useful for these datasets.
Second, we note that the question-answering style datasets (fiqa, hotpotqa, msmarco and nq) seem to benefit the most from an \ac{LLM} approach to query expansion. It is likely that the language model is producing relevant answers towards the query which helps retrieve the relevant passages more effectively.
Across all datasets, the \textbf{\QTDPRF{}} prompt produces the highest average Recall@1K, with the \textbf{\CoT{}} prompt as a close second.

\begin{table*}
    \caption{Recall@1K of various prompts on BEIR using Flan-UL2. \sigsym indicates a statistically significant (paired $t$-test, $p<0.01$) improvement relative to the best classical QE method. The best result per dataset is highlighted in bold.}
    \centering
    \small
    \begin{tabular}{lr|ccc|ccc|ccc|cc}
    \toprule
    & & \multicolumn{3}{c|}{\textbf{Classical QE}} & \multicolumn{8}{c}{\textbf{LLM-based QE}} \\
    Dataset & BM25 & Bo1 & Bo2 & KL & \QTD{} & \QTDZS{} & \QTDPRF{} & \QTE{} & \QTEZS{} & \QTEPRF{} & \CoT{} & \CoTPRF{} \\
    \midrule
    arguana & 98.93 & \textbf{99.00} & \textbf{99.00} & \textbf{99.00} & 98.86 & 98.93 & 98.93 & 98.93 & 98.93 & 98.93 & 98.93 & 98.86 \\
    climate-fever & 46.60 & 45.69 & 45.38 & 45.65 & 47.62 & 47.66 & \textbf{47.94} & 46.08 & 46.44 & 46.44 & 47.42 & 46.81 \\
    cqadupstack & 65.55 & \textbf{66.82} & 66.57 & 66.70 & 65.51 & 64.19 & 65.01 & 65.69 & 65.71 & 65.90 & 66.39 & 66.12 \\
    dbpedia & 63.72 & 64.77 & 64.55 & 64.60 & \textbf{65.89} & 65.47 & 65.78 & 63.55 & 63.92 & 63.93 & 65.77 & 65.06 \\
    fever & 75.73 & 76.28 & 75.83 & 76.32 & \textbf{79.06}\sig & 78.87\sig & 77.29 & 75.78 & 75.79 & 76.27 & 78.21\sig & 77.25 \\
    fiqa & 77.42 & 79.18 & 79.06 & 78.84 & 78.34 & 78.26 & 78.69 & 77.33 & 77.31 & 77.68 & \textbf{80.08} & 79.03 \\
    hotpotqa & 85.78 & 84.84 & 81.71 & 84.65 & 86.90\sig & 85.71 & 87.58\sig & 85.60 & 85.54 & 87.25\sig & 87.54\sig & \textbf{88.79}\sig \\
    msmarco & 73.61 & 75.08 & 75.14 & 74.66 & 76.77 & 75.73 & 78.75 & 73.87 & 73.79 & 74.14 & \textbf{79.58} & 78.36 \\
    nfcorpus & 38.70 & 57.30 & 57.67 & 56.46 & 55.34 & \textbf{59.81} & 59.68 & 43.38 & 44.12 & 47.06 & 52.63 & 53.32 \\
    nq & 78.96 & 81.09 & 80.64 & 80.82 & 85.18\sig & 84.71\sig & 83.53\sig & 79.30 & 79.11 & 80.35 & \textbf{85.46}\sig & 83.11\sig \\
    quora & 99.26 & 99.20 & 99.12 & 99.20 & 99.00 & 98.84 & 98.92 & 99.25 & \textbf{99.29} & 99.26 & 99.17 & 99.21 \\
    scidocs & 57.46 & 59.78 & \textbf{61.03} & 59.86 & 59.09 & 59.78 & 60.10 & 57.88 & 57.70 & 58.32 & 58.51 & 59.69 \\
    scifact & 97.17 & \textbf{97.57} & \textbf{97.57} & \textbf{97.57} & \textbf{97.57} & \textbf{97.57} & \textbf{97.57} & 97.17 & 97.17 & 97.17 & \textbf{97.57} & 97.17 \\
    touche2020 & 84.96 & 85.94 & \textbf{86.38} & 86.01 & 83.61 & 83.44 & 84.54 & 85.21 & 85.02 & 86.04 & 85.51 & 84.58 \\
    trec-covid & 42.58 & 45.21 & \textbf{45.58} & 45.39 & 43.52 & 38.05 & 44.17 & 43.16 & 43.12 & 43.85 & 43.43 & 44.02 \\
    \midrule
    Average & 72.43 & 74.52 & 74.35 & 74.38 & 74.82 & 74.47 & \textbf{75.23} & 72.81 & 72.86 & 73.50 & 75.08 & 74.76 \\
    \bottomrule
    \end{tabular}
    \label{tbl:beir}
\end{table*}

\subsection{The Impact of Model Size}

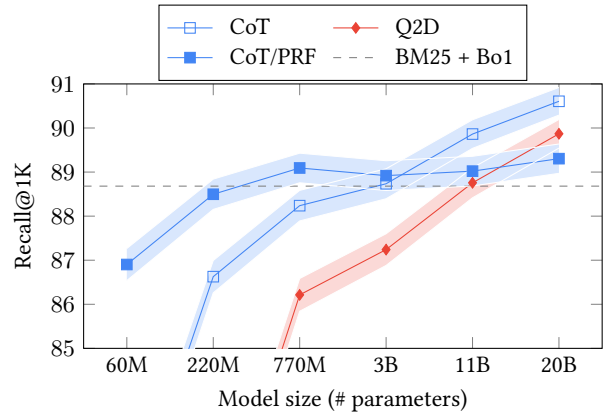
\begin{figure}
    \centering
    \begin{tikzpicture}

\begin{axis}[
    xlabel={Model size (\# parameters)},
    ylabel={Recall@1K},
    xmin=5, xmax=65,
    ymin=85, ymax=91,
    xtick={10,20,30,40,50,60},
    xticklabels={60M,220M,770M,3B,11B,20B},
    ytick={85,...,91},
    legend pos=south east,
    legend columns=2,
    legend style = {
        at={(0.5, 1.3)},
        anchor=north,
        inner sep=3pt,
        style={column sep=0.15cm}
    },
    legend cell align={left},
    height=5.1cm,
    width=\columnwidth
]

\addplot[mark=square,g500blue,name path=cot] plot coordinates {
(10,80.83840128954114)
(20,86.62684275904262)
(30,88.23758202027736)
(40,88.73493109136888)
(50,89.86250555116017)
(60,90.60605515725341)
};
\addlegendentry{\CoT{}}

\addplot[mark=diamond*,g500red,name path=q2d] plot coordinates {
(10,77.39216356964434)
(20,80.17968741009327)
(30,86.21339979005931)
(40,87.24078472646454)
(50,88.75625247770583)
(60,89.86877301905128)
};
\addlegendentry{\QTD{}}

\addplot[mark=square*,g500blue,name path=cotprf] plot coordinates {
(10,86.90255137257678)
(20,88.49652740618946)
(30,89.09286648483334)
(40,88.919266619499)
(50,89.02116545222218)
(60,89.30419111009242)
};
\addlegendentry{\CoTPRF{}}

\addplot[dashed,gray,name path=bo1] plot coordinates {
(5, 88.68086293426597)
(65, 88.68086293426597)
};
\addlegendentry{BM25 + Bo1}

\addplot[g500blue!0,name path=cotL] plot coordinates {
(10,80.41279110480374)
(20,86.25985831901019)
(30,87.890856450472)
(40,88.39472940055467)
(50,89.53797366867956)
(60,90.29234105441712)
};
\addplot[g500blue!0,name path=cotU] plot coordinates {
(10,81.26401147427855)
(20,86.99382719907504)
(30,88.58430759008272)
(40,89.07513278218309)
(50,90.18703743364077)
(60,90.9197692600897)
};
\addplot[g500blue!20] fill between[of=cotL and cotU];
\addplot[g500red!0,name path=q2dL] plot coordinates {
(10,76.93933725683357)
(20,79.74870719777505)
(30,85.84159310910381)
(40,86.88125813752089)
(50,88.41620843257493)
(60,89.54431159117279)
};
\addplot[g500red!0,name path=q2dU] plot coordinates {
(10,77.84498988245512)
(20,80.6106676224115)
(30,86.58520647101481)
(40,87.60031131540819)
(50,89.09629652283671)
(60,90.19323444692976)
};
\addplot[g500red!20] fill between[of=q2dL and q2dU];
\addplot[g500blue!0,name path=cotprfL] plot coordinates {
(10,86.53883427191322)
(20,88.1527303921185)
(30,88.75705788579344)
(40,88.58125674201196)
(50,88.68452587608925)
(60,88.97150525863798)
};
\addplot[g500blue!0,name path=cotprfU] plot coordinates {
(10,87.26626847324036)
(20,88.84032442026043)
(30,89.42867508387322)
(40,89.25727649698605)
(50,89.3578050283551)
(60,89.63687696154688)
};
\addplot[g500blue!20] fill between[of=cotprfL and cotprfU];

\end{axis}

\end{tikzpicture}
    \caption{Performance on MS-MARCO passage ranking dev set across different model sizes. The shaded areas indicate a 99\% confidence interval.}
    \label{fig:modelsizes}
    \Description[A graph showing the performance of different query expansion techniques across various model sizes.]{The x-axis contains the model size (60M up to 20B) and the y-axis contains the Recall@1K ranging from 85 to 91. The BM25+Bo1 baseline is plotted as a horizontal line at 88.68 Recall@1K as it is not affected by model size. The three lines represent three different prompt techniques (labeled CoT, CoT/PRF, and, Q2D) and show that query expansion performance improves as the model size gets larger. At the largest model size, CoT is the best. At a medium-sized model (770M) CoT/PRF is best. CoT is consistently better than Q2D across all model sizes.}
\end{figure}

To understand the practical capabilities and limitations of an \ac{LLM}-based query expander, we compare different model sizes in Figure~\ref{fig:modelsizes}.
We range the model size from 60M parameters (Flan-T5-small) up to 11B (Flan-T5-XXL) and also try a 20B parameter model (Flan-UL2) but note that the latter also has a different pre-training objective.
In general we observe the expected trend that larger models tend to perform better.
The \textbf{\QTD{}} approach requires at least an 11B parameter model to reach parity with the BM25+Bo1 baseline.
In contrast, the \textbf{\CoT{}} approach only needs a 3B parameter model to reach parity.
Furthermore, adding \ac{PRF} documents to the \textbf{\CoT{}} prompt seems to help stabilize the performance for smaller model sizes but does inhibit its performance at larger capacities.
A possible explanation for this behavior is that the \ac{PRF} documents decreases the creativity of the model, as it may focus too much on the provided documents. Although this helps prevent the model from making errors at smaller model sizes, it also inhibits the creative abilities that we wish to leverage at larger model sizes.
The \textbf{\CoTPRF{}} prompt is able to outperform the other prompts at the 770M parameter model size, making it a good candidate for possible deployment in realistic search settings where serving a larger model may be impossible.
Overall, it is clear that large models are able to provide significant gains which may limit the practical application of an \ac{LLM} approach to query expansion. Distillation has been shown to be an effective way to transfer the ability of a large model to a smaller one. We leave the study of distillation of these models for query expansion as future work.

\section{Limitations \& Future Work}
There are a number of limitations in our work:
First, we only study sparse retrieval (BM25) which is where query expansion is important. Dense retrieval systems (e.g. dual encoders) are less prone to the vocabulary gap and, as a result, are less likely to benefit from a query expansion. \citet{wang2023query2doc} has already studied this setting in more detail and we leave the analysis of our prompts for a dense retrieval setting as future work.
Second, our work focuses on Flan~\cite{wei2021finetuned} instruction-finetuned language models. We chose these models due to their ability to follow instructions and the fact that these models are open-source. Our work can naturally be extended to other language models~\cite{brown2020language,chowdhery2022palm,du2022glam,thoppilan2022lamda} and we leave the study of such models as a topic for future research.
Third, we study specific prompt templates (see Appendix~\ref{appendix:prompts}) and there may be other ways to formulate the different prompts.
Finally, the computational cost of \acp{LLM} may be prohibitive to deploy \ac{LLM}-based query expansions in practice. It may be possible to distill the output of the large model into a smaller servable model. How to productionize \ac{LLM}-based query expansions is left as an open problem.
\section{Conclusion}

In this paper we study \ac{LLM}-based query expansions. In contrast to traditional \ac{PRF}-based query expansion, \ac{LLM}s are not restricted to the initial retrieved set of documents and may be able to generate expansion terms not covered by traditional methods. Our proposed method is simple: we prompt a large language model and provide it a query, then we use the model's output to expand the original query with new terms that help during document retrieval.

Our results show that \acl{CoT} prompts are especially promising for query expansion, since they instruct the model to generate verbose explanations that can cover a wide variety of new keywords.
Furthermore, our results indicate that including \ac{PRF} documents in various prompts can improve top-heavy ranking metric performance during the retrieval stage \emph{and} is more robust when used with smaller model sizes, which can help practical deployment of \ac{LLM}-based query expansion.

As demonstrated in this paper, \ac{IR} tasks like query expansion can benefit from \acp{LLM}. As the capabilities of \acp{LLM} continue to improve, it is promising to see their capabilities translate to various \ac{IR} tasks. Furthermore, as \acp{LLM} become more widely available, they will be easier to use and deploy as core parts of \ac{IR} systems which is exciting for both practitioners and researchers of such systems.

\clearpage

\bibliographystyle{ACM-Reference-Format}
\bibliography{references}

\clearpage
\appendix
\section{Prompts}
\label{appendix:prompts}

Table~\ref{tbl:prompts} contains all the prompts tried in this paper. In each prompt \query{} denotes the query for which we want to generate a query expansion. We denote with \querya{}, $\ldots$, \queryd{} the sample queries from the MS-MARCO train set. Similarly, \doca{}, $\ldots$, \docd{} represent relevant passages corresponding to the sampled queries, and, \expa{}, $\ldots$, \expd{} represent corresponding expansions generated with Terrier KL method (at most 20 terms) from those relevant passages. Finally we denote with \prfa{}, $\ldots$, \prfc{} the top 3 retrieved documents using the original query, acting as \acl{PRF} documents. For the \CoT{} prompt, we note that the model tends to output ``The final answer:'' or ``So the final answer is:'' towards the end and we filter those two sentences out prior to concatenating the model output with the query.

\begin{table}[H]
    \centering
    \small
    \renewcommand*\baselinestretch{0.7}\selectfont
    \caption{The various query expansion prompts.}
    \label{tbl:prompts}
    \begin{tabular}[t]{ll}
        \toprule
        \textbf{ID} & \textbf{Prompt} \\
        \midrule 
        \QTD{}~\cite{wang2023query2doc} & \makecell[l]{Write a passage that answers the given query:\\\\
                          Query: \querya{}\\Passage: \doca{}\\\\
                          Query: \queryb{}\\Passage: \docb{}\\\\
                          Query: \queryc{}\\Passage: \docc{}\\\\
                          Query: \queryd{}\\Passage: \docd{}\\\\
                          Query: \query{}\\Passage:} \\
        \midrule
        \QTDZS{} & Write a passage that answers the following query: \query{} \\
        \midrule
        \QTDPRF{} & \makecell[l]{Write a passage that answers the given query based on\\the context:\\\\
                  Context: \prfa{}\\\prfb{}\\\prfc{}\\
                  Query: \query{}\\
                  Passage:} \\
        \midrule
        \QTE{} & \makecell[l]{Write a list of keywords for the given query:\\\\
                          Query: \querya{}\\Keywords: \expa{}\\\\
                          Query: \queryb{}\\Keywords: \expb{}\\\\
                          Query: \queryc{}\\Keywords: \expc{}\\\\
                          Query: \queryd{}\\Keywords: \expd{}\\\\
                          Query: \query{}\\Keywords:} \\
        \midrule
        \QTEZS{} & Write a list of keywords for the following query: \query{} \\
        \midrule
        \QTEPRF{} & \makecell[l]{Write a list of keywords for the given query based on\\the context:\\\\
                  Context: \prfa{}\\\prfb{}\\\prfc{}\\
                  Query: \query{}\\
                  Keywords:} \\
        \midrule
        \CoT{} & \makecell[l]{Answer the following query:\\\\
                           \query{}\\\\
                           Give the rationale before answering} \\
        \midrule
        \CoTPRF{} & \makecell[l]{Answer the following query based on the context:\\\\
                           Context: \prfa{}\\\prfb{}\\\prfc{}\\
                           Query: \query{}\\\\
                           Give the rationale before answering} \\
        \bottomrule 
    \end{tabular}
\end{table}

\vfill
\section{Example Output}
\label{appendix:examples}

Table~\ref{tbl:example:cot} shows the results of a query expansion for both the Flan-T5-Large (770M) model size and the Flan-UL2 (20B) model size. First, note that at the smaller model size, the \CoT{} and \QTD{} prompts are not producing the correct answer which is harmful for retrieval performance. The \CoTPRF{} prompt, being more grounded in its \ac{PRF} documents, avoids this problem and correctly produces the answer ``Tata Motors'' which helps retrieve the relevant passage. At the larger model size (Flan-UL2), all prompts \QTD{}, \CoT{} and \CoTPRF{} produce the correct answer ``Tata Motors''. However, the \CoT{} prompt provides the most verbose explanation towards its answer and has many term overlaps with the relevant passage, improving its overall retrieval performance.

\begin{table}[H]
    \centering
    \caption{Example of query expansions with Flan-UL2. Green highlights the correct answer while blue indicate specific term overlaps that indicate potentially useful query expansions.}
    \label{tbl:example:cot}
    \begin{tabular}{cc|p{5.95cm}}
    \toprule
    & Query & who owns jaguar motors? (ID=1045405) \\
    & Passage & Jaguar is the luxury vehicle brand of \hlc[g100blue]{Jaguar Land Rover}, a \hlc[g100blue]{British multinational car manufacturer} with its headquarters in \hlc[g100blue]{Whitley, Coventry} England, owned by the Indian company \hlc[g100green]{Tata Motors} since 2008. \\
    \midrule
    \multirow{8}{*}{\rotatebox[origin=c]{90}{Flan-T5-Large (770M)}} & \QTD{} & Jaguar Motors is a British luxury car manufacturer based in the United Kingdom. \\
    & \CoT{} & \hlc[g100blue]{Jaguar Land Rover} is the owner of Jaguar. The answer: Jaguar Land Rover. \\
    & \CoTPRF{} & The relevant information is: Jaguar is owned by the Indian automobile manufacturer \hlc[g100green]{Tata Motors} Ltd. So the final answer is the Indian automobile manufacturer \hlc[g100green]{Tata Motors} Ltd. \\
    \midrule
    \multirow{12}{*}{\rotatebox[origin=c]{90}{Flan-UL2 (20B)}} & \QTD{} & Jaguar is a division of \hlc[g100green]{Tata Motors}, a company owned by the Tata family. \\
    & \CoT{} & \hlc[g100blue]{Jaguar Land Rover} is a \hlc[g100blue]{British multinational car manufacturer}, founded by William Lyons in 1931. Its headquarters are in \hlc[g100blue]{Whitley, Coventry}, United Kingdom and is a constituent of the FTSE 250 Index. The company is a wholly owned subsidiary of \hlc[g100green]{Tata Motors} of India. So the final answer is \hlc[g100green]{Tata Motors}. \\
    & \CoTPRF{} & Jaguar is owned by the Indian automobile manufacturer \hlc[g100green]{Tata Motors} Ltd. The final answer: \hlc[g100green]{Tata Motors} Ltd. \\
    \bottomrule
    \end{tabular}
\end{table}

\end{document}